\documentclass{article}

\usepackage{arxiv}

\usepackage[utf8]{inputenc} 
\usepackage[T1]{fontenc}    
\usepackage{hyperref}       
\usepackage{url}            
\usepackage{booktabs}       
\usepackage{amsfonts}       
\usepackage{nicefrac}       
\usepackage{microtype}      
\usepackage{lipsum}
\usepackage{spconf,amsmath,graphicx}
\usepackage{amsfonts,amssymb}
\usepackage{siunitx}
\usepackage{subfigure}
\usepackage{caption}
\usepackage{flushend}
\usepackage{fancyhdr}

\title{SAR image autofocusing using Wirtinger Calculus and Cauchy regularization}

\name{Zi-Yao Zhang, Odysseas Pappas, Alin Achim\thanks{This work was supported in part by the EPSRC under grant EP/R009260/1 (AssenSAR), in part by a Royal Society Enhancement Award (RGF-EA-182086) and in part by a Chinese PhD Scholarship Council studentship (to Zhang).}}
\address{Visual Information Lab, University of Bristol, UK}

\begin{document}
\maketitle

\begin{abstract}
In this paper, an optimization model using Cauchy regularization is proposed for simultaneous SAR image reconstruction and autofocusing. A coordinate descent framework in which the desired image and the phase errors are optimized alternatively is designed to solve the model. For the subproblem of estimating the image, we utilize the techniques of Wirtinger calculus to directly minimize the cost function which involves complex variables. We also utilise a state-of-the-art, sparsity-enforcing Cauchy regularizer. The proposed method is demonstrated to give impressive autofocusing results by conducting experiments on both simulated scene and real SAR image.
\end{abstract}


\section{Introduction}
Synthetic aperture radar (SAR) is capable of acquiring data in all kinds of weather and lighting conditions. As a coherent imaging radar system, it is often mounted on an airplane or satellite platform, where it transmits a frequency modulated signal to a scene on the ground and records the measured response in flight. A detailed introduction about the working mechanisms of SAR can be found in \cite{moreira2013tutorial, chan2008introduction, ouchi2013recent}.
However, this process of data acquisition is often plagued by phase errors. Due to inaccuracies in the SAR platform trajectory measurement, as well as due to possible movement of objects in the scene, the observed data (radar returns) will contain phase errors. These in turn lead to a defocusing effect in the formed SAR image. Techniques aiming to directly estimate these phase errors from the raw SAR data and then remove them so as to improve formed image quality are called autofocusing techniques.

Among the earliest autofocusing techniques, phase gradient autofocus (PGA) \cite{wahl1994phase} is a very well-known method  which computes the phase error according to the estimation of its gradient. Mapdrift autofocus is another classical technique for autofocusing \cite{calloway1994subaperture} which estimates phase errors from the maps reconstructed for each sub-aperture.

Differing from those approaches are many optimization-based methods proposed in recent years. These mainly belong to one of two categories. The first maximizes a sharpness metric in the form of a power function \cite{fienup2003aberration, morrison2007sar}, or image entropy  \cite{kragh2006monotonic,kantor2017minimum}. The second adopts an inverse problem approach. Based on a forward observation model relating the corrupted phase history to the underlying SAR image, SAR autofocusing is formulated as an inverse problem, and various regularized variational models have been proposed to obtain its solution. For instance, Onhon et al. uses the $p$th power of the approximate $l_{p}$ norm as the regularization term and a coordinate descent framework to solve the problem \cite{onhon2011sparsity}. There are also various methods addressing the problem in a compressive sensing context, such as the majorization-minimization based method \cite{kelly2012auto, kelly2014sparsity}, iteratively re-weighted augmented Lagrangian based method \cite{gungor2015augmented,  gungor2017autofocused} and conjugate gradient based method with two different regularization terms (approximate $l_{1}$ and approximate total variation regularization) \cite{ugur2012sar}. 

Besides these, there are also some methods which are built on traditional SAR imaging methods. An autofocusing method which maximizes a sharpness metric for each pulse in the imaging process of back-projection is proposed in \cite{ash2011autofocus} and further extended for the case involving moving ship targets \cite{sommer2019backprojection}. A polar format algorithm based autofocusing approach \cite{kantor2019polar} which combines \cite{onhon2011sparsity} with classical autofocusing method like PGA has recently been proposed as well.

In this paper, we formulate the SAR autofocusing problem as an inverse problem, and adopt a coordinate descent framework to jointly estimate the desired SAR image and the phase error. The optimization process of the phase error is the same as that in \cite{onhon2011sparsity}. However, for the optimization of the desired SAR image, we use Cauchy regularization on the magnitude of the desired image (thus we call it "magnitude Cauchy") and propose an efficient method based on Wirtinger calculus  \cite{brandwood1983complex,  kreutz2009complex} to handle the complex nature of this subproblem. 

The rest of this paper is organized as follows. In Section 2, the data acquisition model for SAR will be introduced briefly. In Section 3, our proposed methods is described. In Section 4, experimental results are shown to demonstrate the effectiveness of our methods. Finally, conclusions are drawn in section 5.

\section{Data acquisition model}

In this paper, we consider a SAR platform operating in spotlight mode, whose transmitted signal at each azimuth position can be formulated as:
\begin{equation} \label{eq1}
	s(t)=Re\{e^{j(\omega_{0}t+\alpha t^{2})}\},
\end{equation}
where $\omega_{0}$ is the carrier frequency and 2$\alpha$ is the chirp rate. The relationship between the observed data $r_{m}(t)$ at the $m$th aperture position and the underlying SAR image $F(x,y)$ can be described by:
\begin{equation} \label{eq2}
	r_{m}(t)=\iint F(x,y)e^{-jU(xcos\theta+ysin\theta)}dxdy.
\end{equation}
The region over which the integral is computed is $x^{2}+y^{2}\le L^{2}$, with $L$ being the radius of the circular patch on the ground to be imaged. $\theta$ is the look angle, and $U$ is defined by
\begin{equation} \label{eq3}
	U=\frac{2}{c}(\omega_{0}+2\alpha(t-\tau_{0})),
\end{equation}
with $\tau_{0}$ being the demodulation time. The discretized version for this model is
\begin{equation} \label{eq4}
	r_{m}=C_{m}f,
\end{equation}
where $r_{m}$ and $C_{m}$ are the vector form of the phase history and the observation matrix for the $m$th aperture position resptectively. $f$ is the vector form of the underlying SAR image. Stacking (4) for all the aperture positions, and considering phase error as well as noise, the model becomes
\begin{equation} \label{eq5}
	g=C(\phi)f+n,
\end{equation}
with $g$ being the corrupted phase history, $\phi$ being the phase error, and $n$ being Gaussian white noise. $C(\phi)$ is the corrupted observation matrix. Taking the case of 1D phase error varying along azimuth direction as an example, we have
\begin{equation} \label{eq6}
	C_{m}(\phi)=e^{j\phi_{m}}C_{m},
\end{equation}
where $C_{m}(\phi)$ and $\phi_{m}$ are the corrupted observation matrix and phase error for the $m$th aperture position respectively.

\section{Wirtinger Coordinate Descent Autofocusing}
\label{sec:pagestyle}
We propose to minimize the following cost function so that the desired SAR image $f$ and the phase error $\phi$ are jointly estimated, and therefore SAR image reconstruction and phase error removal are achieved simultaneously:
\begin{equation} \label{eq7}
	J(f,\phi)=\|g-C(\phi)f\|_{2}^{2}-\lambda \sum_{i=1}^{N}\ln{\frac{\gamma}{\gamma^{2}+|f_{i}|^{2}}}.
\end{equation}
$\lambda$ is the regularization parameter and $\gamma$ is the scale parameter for Cauchy distribution. 

The penalty term used in (7) is a Cauchy regularization merely imposed on the magnitude of the latent SAR image and we therefore refer to it as  "magnitude Cauchy regularization". Like the $l_{p}$ norm, it too is a sparsity enforcing regularization term, and its effectiveness has already been demonstrated in SAR imaging and other inverse problems \cite{karakucs2020solving, karakus2020convergence}.

Since we want to solve $f$ and $\phi$ jointly, we adopt a coordinate descent autofocusing framework similar to \cite{onhon2011sparsity}. Specifically, $f$ and $\phi$ are updated alternatively by fixing one of them while optimizing the other. This iterative process will terminate when the relative error between $f^{(n)}$ and $f^{(n+1)}$ is smaller than $10^{-3}$. 

\subsection{Image reconstruction}
We first discuss the iterative estimation of the desired image $f$. In each iteration of the proposed Wirtinger coordinate descent autofocusing (WCDA) framework, the subproblem to be solved is
\begin{equation} \label{eq8}
	f^{(n+1)}=\mathrm{arg}\min_{f\in \mathbb{C}^{(n)}}\|g-C(\phi^{(n)})f\|_{2}^{2}-\lambda \sum_{i=1}^{N}\ln{\frac{\gamma}{\gamma^{2}+|f_{i}|^{2}}}.
\end{equation}
 
Unlike many other imaging-related inverse problem formulations, this is an optimization problem involving a complex unknown vector, and needs to be handled with the appropriate mathematical tools. Specifically, the derivative results for real optimization problems need to be modified or redefined to fit the complex case. Therefore, we here present a technique for directly computing the complex gradient and solving this problem using Wirtinger calculus. Wirtinger calculus is a powerful theory covering the analysis of real-valued functions of complex variables, and thus is applicable for this subproblem. It is also a rather elegant approach, because it can address the problem in a more concise way, i. e., there is no need to expand the complex variables as their real counterparts in the computational process.

To find the minimum of (8), we compute the gradient of it directly using Wirtinger calculus. The definition of the gradient for a real-valued function $h(f)$ with a complex vector variable $f$ is first proposed in \cite{brandwood1983complex} and further extended in \cite{kreutz2009complex}, and is as follows:
\begin{equation} \label{eq9}
	\nabla_{f}h=\Omega_{f}^{-1}(\frac{\partial h}{\partial f})^{H}.
\end{equation}
where $\Omega_{f}^{-1}$ is a metric tensor. In this paper we use Brandwood’s setting, i. e., let it be the identity matrix, then we have
\begin{equation} \label{eq10}
	\nabla_{f}h=(\frac{\partial h}{\partial f})^{H}.
\end{equation}
Since the cost function is real-valued, according to \cite{kreutz2009complex}, we have:
\begin{equation} \label{eq11}
	(\frac{\partial h}{\partial f})^{H}=\overline{(\frac{\partial h}{\partial f})}^{T}=(\frac{\partial h}{\partial \overline{f}})^{T}=(\frac{\partial h}{\partial \overline{{f}_{1}}},...,\frac{\partial h}{\partial \overline{{f}_{N}}})^{T}.
\end{equation}

Each component of the last term in the right side of (11) can be computed using the chain rule and the definition of the conjugate $\mathbb{R}$-derivative \cite{kreutz2009complex}. As a result, for the second term of (8), if we denote
\begin{equation} \label{eq12}
	R(f)=-\sum_{i=1}^{N}\ln{\frac{\gamma}{\gamma^{2}+|f_{i}|^{2}}},
\end{equation}
then we have
\begin{equation} \label{eq13}
	(\nabla R(f))_{i}=\frac{f_{i}}{\gamma^{2}+|f_{i}|^{2}}. i=1,...,N.
\end{equation}

For the first term of (8), when we rewrite the square of $l_{2}$ norm as an inner product, and use the derived result in \cite{brandwood1983complex}, it is obvious that:
\begin{equation} \label{eq14}
\nabla \|g-C(\phi^{(n)})f\|_{2}^{2}=C(\phi^{(n)})^{H}(C(\phi^{(n)})f-g).
\end{equation}

Therefore, the gradient of (8) can be written as:
\begin{equation} \label{eq15}
\nabla J(f,\phi)=C(\phi^{(n)})^{H}(C(\phi^{(n)})f-g)+\lambda W(f)f,
\end{equation}
where 
\begin{equation} \label{eq16}
W(f)=diag(s),
\end{equation}
\begin{equation} \label{eq17}
s_{i}=\frac{1}{\gamma^{2}+|f_{i}|^{2}}, i=1,...,N.
\end{equation}

Now we set (15) to zero, because this is the necessary and sufficient condition for a stationary point of a real-valued complex function \cite{brandwood1983complex, kreutz2009complex}. This leads to
\begin{equation} \label{eq18}
	[C(\phi^{(n)})^{H}C(\phi^{(n)})+\lambda W(f)]f=C(\phi^{(n)})^{H}g.
\end{equation}
It is worth pointing out that the solution found in this way is not necessarily the global minimum due to the non-convexity of the Cauchy penalty, but the experimental results in Section 4  imply that the obtained solution is sufficiently good.

We can now rewrite (18) as $Af=b$, where $b=C(\phi^{(n)})^{H}g$, $A=C(\phi^{(n)})^{H}C(\phi^{(n)})+\lambda W(f)$. 
Since $W(f)$ depends on $f$, so does $A$. This makes (18) nonlinear in $f$, and it is difficult to find a closed form solution for it. However, if we approximate the $f$ in $W(f)$ with the $f$ computed by the last iteration of the coordinate descent framework, $A$ can be turned into a constant matrix, and thus $Af=b$ becomes a linear system of equations. 
That is to say, when computing an unknown $f^{(n+1)}$, we actually solve
\begin{equation} \label{eq19}
	[C(\phi^{(n)})^{H}C(\phi^{(n)})+\lambda W(f^{(n)})]f^{(n+1)}=C(\phi^{(n)})^{H}g.
\end{equation}
The solution for this equation can be efficiently obtained by using the conjugate gradient algorithm \cite{barrett1994templates}.

\subsection{Phase Error Optimization}

For 1D phase error varying along the azimuth direction, after obtaining $f^{(n+1)}$, for the aperture position $m$, the phase error can be updated by solving:

\begin{equation} \label{eq20}
	\phi_{m}^{(n+1)}=\mathrm{arg}\min_{\phi_{m} }\|g_{m}-e^{j\phi_{m}}C_{m}f^{(n+1)}\|_{2}^{2},
\end{equation}
with $g_{m}$ and $C_{m}$ being the parts of $g$ and $C$ corresponding to the $m$th aperture position. According to \cite{onhon2011sparsity}, the solution is 
\begin{equation} \label{eq21}
	\phi_{m}^{(n+1)}=-\arctan(-\frac{\text{Re}{[f^{(n+1)}]^{H}C_{m}g_{m}}}{\text{Im}{[f^{(n+1)}]^{H}C_{m}g_{m}}}).
\end{equation}
This estimation is then used to update the corrupted observation matrix as:
\begin{equation} \label{eq22}
	C_{m}(\phi_{m}^{(n+1)})=e^{j\phi_{m}^{(n+1)}}C_{m}.
\end{equation}

The cases of 2D phase errors varying in both range direction and cross-range direction can also be solved by similar methods, see \cite{onhon2011sparsity} for more details.

\section{Experimental results}
\label{sec:typestyle}

In this paper, the same radar system model as in \cite{onhon2011sparsity} is used, and its parameters are listed in Table \ref{tab:radarparameters}.

\begin{table}[!htbp]
	\small
	\caption{\textbf{Parameters of the radar system.}}
	\begin{center}
		\begin{tabular}{|l|c|c|}
			\hline
			\textbf{Carrier Frequency}&$2\pi \times 10^{10} rad/s$\\\hline
			\textbf{Chirp Rate}&$2\pi \times 10^{12} rad/s^{2}$\\\hline
			\textbf{Pulse Duration}&$4 \times 10^{-4}s$\\\hline
			\textbf{Angular Range}&$\ang{2.3}$\\\hline
		\end{tabular}
	\end{center}
	\label{tab:radarparameters}
\end{table}

In each experiment, this radar system model is used to generate a simulated phase history from a given scene. The corrupted phase history is then obtained by adding 1D random phase error along the azimuth direction as well as white Gaussian noise to the originaly simulated phase history. This corrupted phase history is then used to reconstruct a SAR image while correcting for the phase error.

The proposed method is compared to two methods. One is the sparsity driven autofocus (SDA) method \cite{onhon2011sparsity} which is a state of the art autofocusing technique operating in an inverse problem framework similar to the one employed here, while the other is the traditional polar format algorithm \cite{walker1980range} which doesn't involve autofocusing.

Apart from visual comparison, two numerical metrics are also computed to better assess the performance of each method. They are the mean square error between the reconstructed SAR image and the original scene, and the entropy of the reconstructed SAR image. For both of these metrics, smaller values indicate better performance.

In the first experiment, we use a simulated scene measuring $32\times 32$ pixels. The visual results are presented in Fig.1, and the numerical results are listed in Table 2. It can be observed that the visual result of polar format suffers from severe defocusing effect. However, the reconstructed images by SDA and the proposed method are both very focused and highly resemble the original scene. The values of MSE and entropy for the proposed method are lower, suggesting that its result is sharper and more similar to the original scene.

In the second experiment, a real TerraSAR-X image  serves as the scene. Due to the high computational burden of our method for scenes of large size, a $64\times 64$ patch is cut from the original image and regarded as an input scene. The corrupted pseudo-phase history is generated from it as explained above. 

Fig. 2 shows the reconstructed images by all 3 methods, while Table 2 contains the corresponding numerical results. According to Fig. 2, the polar format algorithm once again gives reconstructed results with smeared targets. However, both SDA and the proposed method can remove phase errors effectively and present focused targets, displaying significant improvement over the result of the polar format algorithm. Nevertheless, the results of the numerical indices in Table 2 demonstrate that the proposed method outperforms SDA. 

\begin{figure}[t]
	\renewcommand{\figurename}{\textbf{Fig.}}
	\centering
	\subfigure[]{
		\includegraphics[width=1.5in]{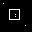}}
	\subfigure[]{
		\includegraphics[width=1.5in]{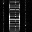}}
	\subfigure[]{
		\includegraphics[width=1.5in]{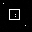}}
	\subfigure[]{
		\includegraphics[width=1.5in]{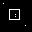}}
	\captionsetup{font={normalsize}}
	\caption{Visual results for Scene 1, a simulated $32\times 32$ scene obtained by various methods. (a) original scene, (b) polar format reconstruction, (c) SDA, (d) the proposed method.}
\end{figure} 

\begin{figure}[t]
	\renewcommand{\figurename}{\textbf{Fig.}}
	\centering
	\subfigure[]{
		\includegraphics[width=1.5in]{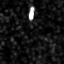}}
	\subfigure[]{
		\includegraphics[width=1.5in]{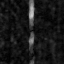}}
	\subfigure[]{
		\includegraphics[width=1.5in]{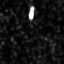}}
	\subfigure[]{
		\includegraphics[width=1.5in]{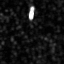}}
	\captionsetup{font={normalsize}}
	\caption{Visual results for Scene 2, a $64\times 64$ patch from TerraSAR-X obtained by various methods. (a) original scene, (b) polar format reconstruction, (c) SDA, (d) the proposed method.}
\end{figure}  



\begin{table}[!t]
	\small
	\caption{\textbf{Numerical metric evaluation of the experimental results for SDA and the proposed method.}}
	\begin{center}
		\begin{tabular}{|l|c|c|}
			\hline
			\multicolumn{3}{|c|}{\textbf{Scene 1}} \\ \hline  
			& \textbf{MSE}&\textbf{Entropy}\\\hline
			\textbf{SDA}&5.4310$\times  10^{-6}$&1.4621\\ \hline
			\textbf{Proposed}&1.2227$\times  10^{-6}$&0.3327\\ \hline
			\hline
			\multicolumn{3}{|c|}{\textbf{Scene 2}} \\ \hline
			& \textbf{MSE}&\textbf{Entropy}\\ \hline
			\textbf{SDA}& 6.4964$\times  10^{-5}$ &5.4410\\ \hline
			\textbf{Proposed}& 6.3029$\times  10^{-5}$ &5.4333\\ \hline
			
		\end{tabular}
	\end{center}
	\label{tab:allres}
\end{table}

\section{Conclusions}
\label{sec:typestyle}

In this paper, an optimization model regularized by Cauchy penalty is proposed to simultaneously reconstruct a SAR image and achieve autofocusing. A coordinate descent framework is adopted to solve this inverse problem, in which Wirtinger calculus is utilized to directly solve one of the subproblems involving complex optimization. Experimental results on simulated phase histories derived from a simulated scene and a real SAR image demonstrate that the proposed method can reconstruct a highly focused image and effectively remove phase erro

\bibliographystyle{unsrt}  
\bibliography{references}  

\end{document}